\documentclass{article}
\pdfoutput=1

\usepackage{arxiv}

\usepackage[utf8]{inputenc} 
\usepackage[T1]{fontenc}    
\usepackage{hyperref}       
\usepackage{url}            
\usepackage{booktabs}       
\usepackage{amsfonts}       
\usepackage{nicefrac}       
\usepackage{microtype}      
\usepackage{lipsum}		
\usepackage{graphicx}
\usepackage[numbers]{natbib}
\usepackage{doi}

\usepackage{subcaption}

\title{Music Emotion Prediction \\Using Recurrent Neural Networks}

\author{ \href{https://xiyahc.github.io/about/}{\includegraphics[scale=0.06]{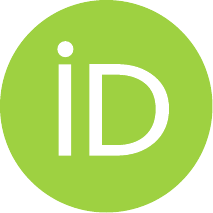}\hspace{1mm}Xinyu Chang}\\
	Johns Hopkins University\\
	\texttt{xchang23@jh.edu} \\
	\And
	{\includegraphics[scale=0.06]{orcid.pdf}\hspace{1mm}Xiangyu Zhang} \\
	Johns Hopkins University\\
	\texttt{xzhan344@jh.edu} \\
	\AND
	{\includegraphics[scale=0.06]{orcid.pdf}\hspace{1mm}Haoruo Zhang} \\
	Johns Hopkins University\\
	\texttt{hzhan237@jh.edu} \\
	\And
	{\includegraphics[scale=0.06]{orcid.pdf}\hspace{1mm}Yulu Ran} \\
	Johns Hopkins University\\
	\texttt{yran2@jh.edu} \\
}

\date{}

\renewcommand{\headeright}{}
\renewcommand{\undertitle}{}
\renewcommand{\shorttitle}{}

\hypersetup{
pdftitle={A template for the arxiv style},
pdfsubject={q-bio.NC, q-bio.QM},
pdfauthor={David S.~Hippocampus, Elias D.~Striatum},
pdfkeywords={First keyword, Second keyword, More},
}

\begin{document}
\maketitle

\begin{abstract}
	This study explores the application of recurrent neural networks to recognize emotions conveyed in music, aiming to enhance music recommendation systems and support therapeutic interventions by tailoring music to fit listeners' emotional states. We utilize Russell's Emotion Quadrant to categorize music into four distinct emotional regions and develop models capable of accurately predicting these categories. Our approach involves extracting a comprehensive set of audio features using Librosa and applying various recurrent neural network architectures, including standard RNNs, Bidirectional RNNs, and Long Short-Term Memory (LSTM) networks. Initial experiments are conducted using a dataset of 900 audio clips, labeled according to the emotional quadrants. We compare the performance of our neural network models against a set of baseline classifiers and analyze their effectiveness in capturing the temporal dynamics inherent in musical expression. The results indicate that simpler RNN architectures may perform comparably or even superiorly to more complex models, particularly in smaller datasets. We've also applied the following experiments on larger datasets: one is augmented based on our original dataset, and the other is from other sources. This research not only enhances our understanding of the emotional impact of music but also demonstrates the potential of neural networks in creating more personalized and emotionally resonant music recommendation and therapy systems.
\end{abstract}


\section{Introduction}
Our project aims to leverage neural networks to analyze the emotional expressions conveyed by music. Consider the profound impact music has on our emotions, such as evoking feelings of happiness or sadness. Our objective is to employ these techniques to enhance music recommendation systems, tailoring selections more closely to listeners' moods. Additionally, we anticipate that our findings could support therapeutic interventions, where specific musical moods are used to aid individuals dealing with emotional challenges. Our methodology involves developing robust techniques for labeling and predicting the emotional categories of music, based on Russell's Emotion Quadrant. Through this research, we seek to establish a more accurate framework for classifying the emotional content of arbitrary audio samples into specific emotional regions. This endeavor not only advances our understanding of music's emotional dimensions but also enhances the personalization of music therapy and recommendation systems.
\\
\\

\section{Relative Works}
\label{sec:headings}

\begin{center}
Novel Audio Features for Music Emotion Recognition \cite{panda2018novel}
\end{center}
In this paper, Renato Panda and colleagues introduce a set of novel audio features specifically designed for music emotion recognition. These features encompass melodic, dynamic, rhythmic, musical texture, and expressivity aspects. The authors elaborate on the methods used to extract these features from audio clips and describe the algorithms they developed for this process. Additionally, they provide well-managed datasets, which we have chosen to utilize in our project.

\begin{center}
A Novel Music Emotion Recognition Model Using Neural Network Technology \cite{yang2021novel} 
\end{center}
In this paper, Jing Yang discusses the limitations of neural networks, particularly the ease with which they can fall into local solutions and their low efficiency due to backpropagation. To address these issues, Yang introduces the Artificial Bee Colony Algorithm as a potential solution. The paper presents the idea of using neural network models for music emotion recognition tasks and explores the possibility of incorporating the Artificial Bee Colony Algorithm to enhance the performance of these models, which we have chosen to attempt in our project.

\section{Method}
\label{sec:headings}

\subsection{Datasets}

We used 3 datasets to test our model: 4Q audio emotion dataset (small size), the augmented 4Q audio emotion dataset  (medium size), and the MTG-Jamendo Dataset (large size). 

\subsubsection{4Q audio emotion dataset}

The dataset contains 900 audio clips,  gathered from AllMusic API. According to Russell’s model\cite{panda2018musical}, the datasets are annotated into 4 quadrants. 

\subsubsection{Augmented dataset}
To generate syntactic data for audio, we applied noise injection, shifting time, changing pitch, and speed. After data augmentation, we have 3600 audio clips in total. 

By using Librosa, we computed the chromagram of each audio clip and converted all 3600 30-sec audio clips into 14 features ['chroma\_stft', 'chroma\_cqt', 'chroma\_vqt', 'melspectrogram', 'mfcc', 'rms', 'spectral\_centroid', 'spectral\_bandwidth', 'spectral\_contrast', 'spectral\_flatness', 'spectral\_rolloff', 'tonnetz', 'zero\_crossing\_rate', 'tempo']. To take advantage of Recurrent Neural Networks (RNNs) to process time series data, we stacked 14 features into a single feature. Therefore, after processing, X is in the shape of (3600, 204, 1295), while y comprises four labels and exhibits a shape of (3600, ). 

\subsubsection{MTG-Jamendo Dataset}

The dataset contains over 55,000 full audio tracks with 195 tags from genre, instrument, and mood/theme categories. We test our models and baseline model on subsets of the MTG-Jamendo Dataset\cite{bogdanov2019mtg}, which contains nearly 14,000 audio clips and 59 tags.

\subsection{Data Visualization}



\begin{figure}[h]
\centering
\includegraphics[width=8cm]{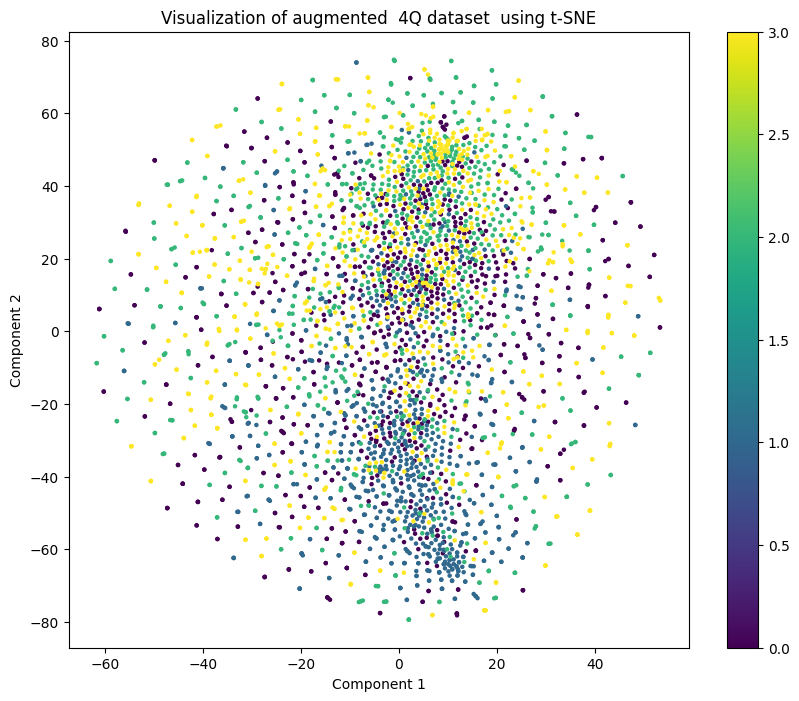}
\caption{Data Visualization for Augmented 4Q Dataset}
\label{visualization}
\end{figure}

 The following is a t-SNE plot obtained by embedding 14 features into 2 dimensions, which presents a good visual representation of how the features are distributed. From Figure \ref{visualization} we can find that label 2 (green), and label 3 (yellow) points tend to be concentrated in the upper-middle right position, and label 1 (light blue) in the lower-middle right position. Although the clustering is not very tight, points of certain classes seem to tend to cluster together, suggesting that there may be some inherent structure in the multidimensional space.  

\section{Experiments}

\subsection{Hypothesis}

Audio clips contain a lot of temporal information. Models based on recurrent neural networks can connect previous and subsequent time steps. Therefore, we tested Recurrent Neural Networks (RNNs), Bidirectional Recurrent Neural Networks (BRNNs), and Long Short-Term Memory (LSTM) on our dataset. We hypothesize that by extracting 14 spectral and rhythmic features, RNNs-related models(RNNs, BRNNs, LSTM) can perform better than traditional Machine Learning models to classify audio clips into 4 Russell’s Emotion Quadrants.

\subsection{Baseline Models}

The baseline classifiers we select encompass a diverse set of machine-learning algorithms that include both traditional statistical models and more advanced machine-learning techniques. This collection comprises Logistic Regression (LR), RidgeClassifier (RC), Linear Support Vector Classification (LSVM), and Support Vector Machine (SVM), which are commonly used for linear and non-linear decision boundary classification. Additionally, the ensemble-based Random Forest Classifier (RFC) and AdaBoost Classifier (ABC) offer improved predictive performance through the aggregation of multiple learners. Gaussian Naive Bayes (GNB), Linear Discriminant Analysis (LDA), and Quadratic Discriminant Analysis (QDA) apply probabilistic frameworks to classify data based on statistical distributions. One-vs-Rest Classifier (OvRC) simplifies multi-class problems into multiple binary classifications. The K-Neighbors Classifier (KNC) utilizes neighborhood-based learning, while the Multilayer Perceptron (MLP) represents a deep learning approach for more complex patterns. We suppose these classifiers provide a comprehensive and varied set of benchmarks for comparative analysis.

\begin{figure}[h]
\centering
\includegraphics[width=15cm]{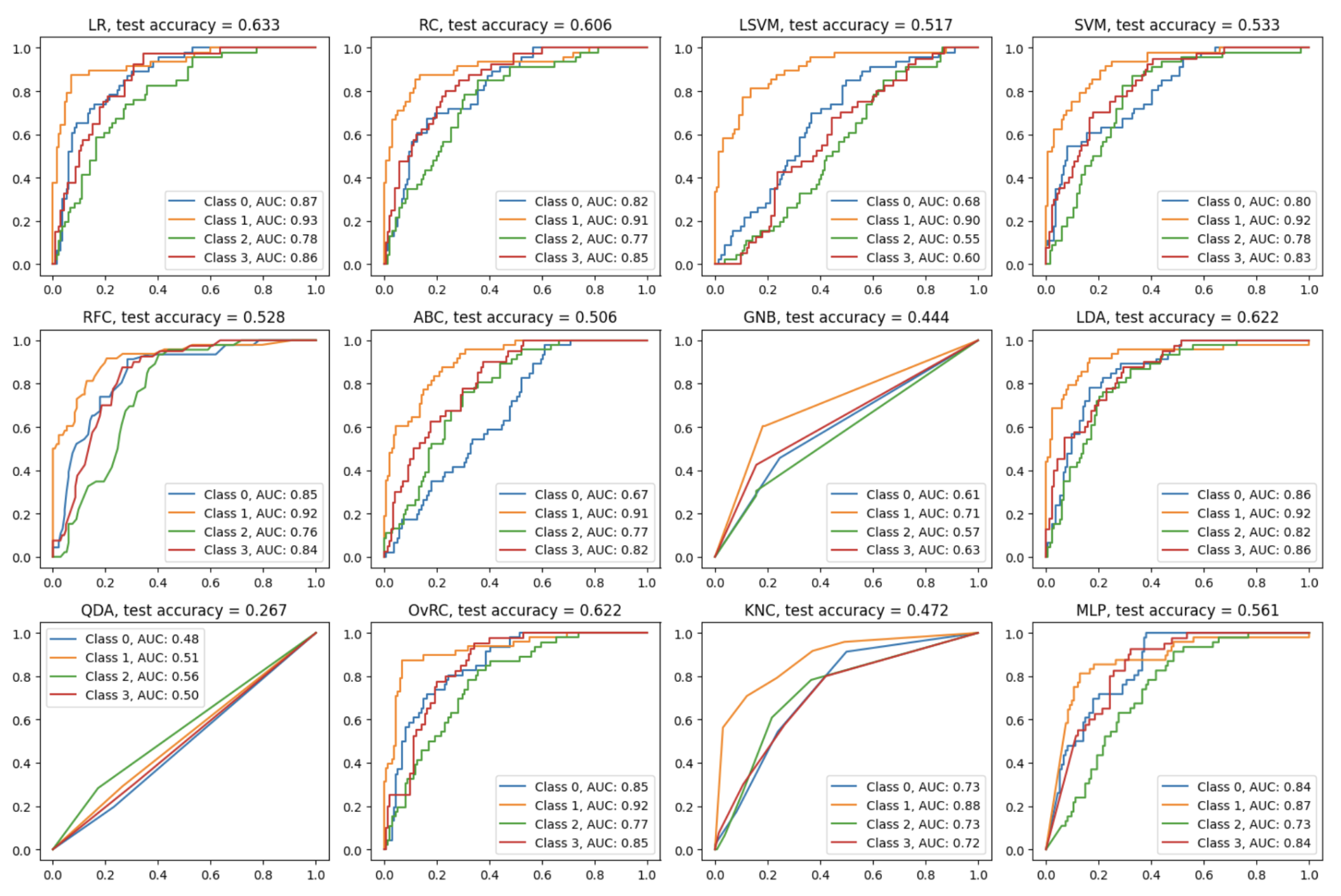}
\caption{Plots of Test Accuracy of Baseline Models}
\label{BaselinePlot1}
\end{figure}

From the result of 12 scikit-learn classifiers shown in Figure \ref{BaselinePlot1}, the top three best models are Logistic Regression,  Linear Discriminant Analysis, and One-vs-Rest Classifier with accuracy 0.633, 0.622, 0.622 respectively. 

\subsection{RNNs Models}

Recurrent Neural Networks(RNNs) are specialized for processing sequential data $x(t)=x(1), …, x(\tau)$, where $t$ represents the time step index ranging from 1 to $\tau$. In this paper, we will explore three specific RNN models: standard RNNs, Bidirectional RNNs(BRNNs), and Long Short-Term Memory(LSTM), which is a specialized variant of RNNs. We will discuss these models in greater detail in the following sections.

While implementing the code of these models, we will use the AdamW optimizer instead of Adam due to its improved implementation of weight decay. Additionally, we will compute cross-entropy loss, a prevalent method for classification tasks which is effective for calculating loss when outputs are probabilities. For RNNs and BRNNs, gradient clipping is applied to prevent gradient explosion or vanishing, whereas LSTM already prevents such issues as a highlight for its model, therefore gradient clipping is not applied in LSTM. Early-stopping is not used in this experiment, instead, we manually choose the iteration number(epoch number) to stop where the evaluation accuracy won't increase after that stop point.

\subsubsection{RNNs}

RNNs possess a form of “memory” that captures information from previous computations. Essentially, an RNN can be un-rolled to multiple copies of the same network, where each instance passes information to its successor.

\begin{figure}[h]
\centering
\includegraphics[width=13cm]{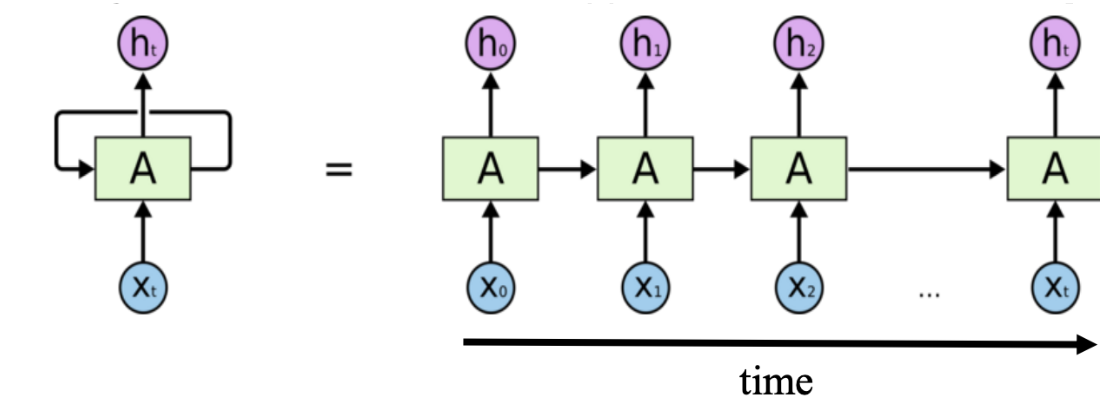}
\caption{Unrolled RNNs Models}
\label{RNN}
\end{figure}

Figure \ref{RNN} displays RNNs after unrolling. We will use numerical features extracted from 900 audio clips using Librosa as inputs for these RNNs. Since the basic structures are similar, these inputs will also be used for the other two RNNs models we plan to explore.

\textbf{Model Input}: After data preparation, the 14 features of our original dataset, transformed using the Librosa package, will be organized into a vector of shape (204, 1295) for each of the 900 samples, where 1295 represents the number of time frames analyzed by Librosa. Consequently, our dataset will have the shape ([900, 204, 1295]), and the labels will be shaped ([900]), containing values {0, 1, 2, 3}. These values represent the 1st, 2nd, 3rd, and 4th quadrants of Russell's Emotion Quadrant, respectively.

\textbf{Model Output}: The output will be one of the {0, 1, 2, 3} which is predicted by the model.

\textbf{Model Architecture}: We used Dropout with probability 0.2 and RELU as the activation function within the fully connected layer. Since we are using many-to-one RNNs, The hidden state of the last time step will be passed to the fully connected layer instead of the output of the model to get the final output.

\begin{figure}[h]
\centering
\includegraphics[width=10cm]{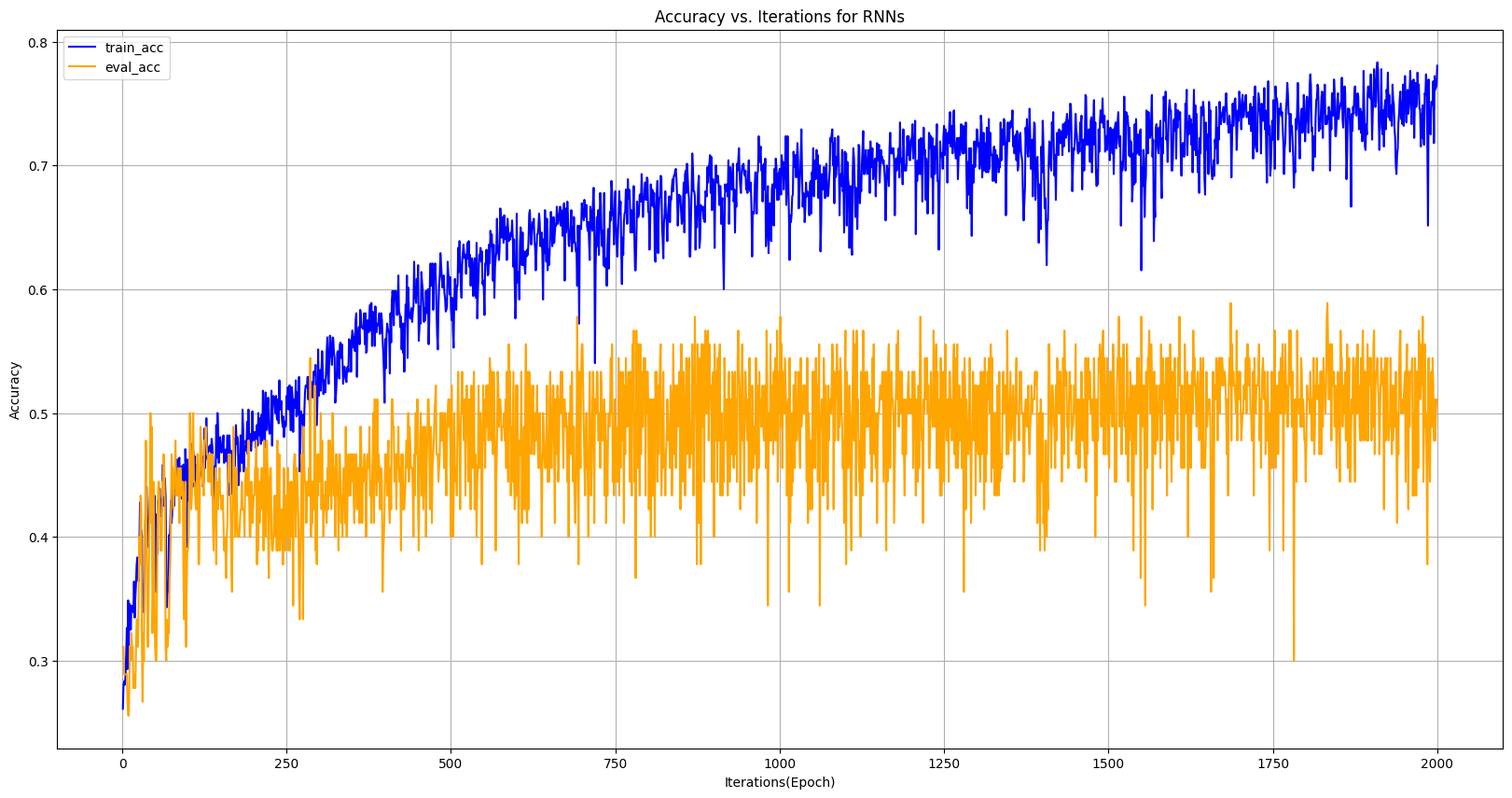}
\caption{Train and Evaluation Accuracy vs. Iterations for RNNs}
\label{RNNTestAcc}
\end{figure}

\textbf{Train}: For the basic RNNs model, we run 2000 epochs and get the train accuracy and evaluation accuracy shown in Figure \ref{RNNTestAcc}.

We pick the epoch numbers to stop by viewing if the evaluation accuracy starts to decrease on a large scale, or if the training accuracy stops increasing.

\subsubsection{BRNNs}

Bidirectional Recurrent Neural Networks (BRNNs) can connect to both previous and subsequent time steps, unlike basic RNNs, which only link to previous time steps. We plan to explore whether this model offers an improvement over the traditional RNNs.

\begin{figure}[h]
\centering
\includegraphics[width=3cm]{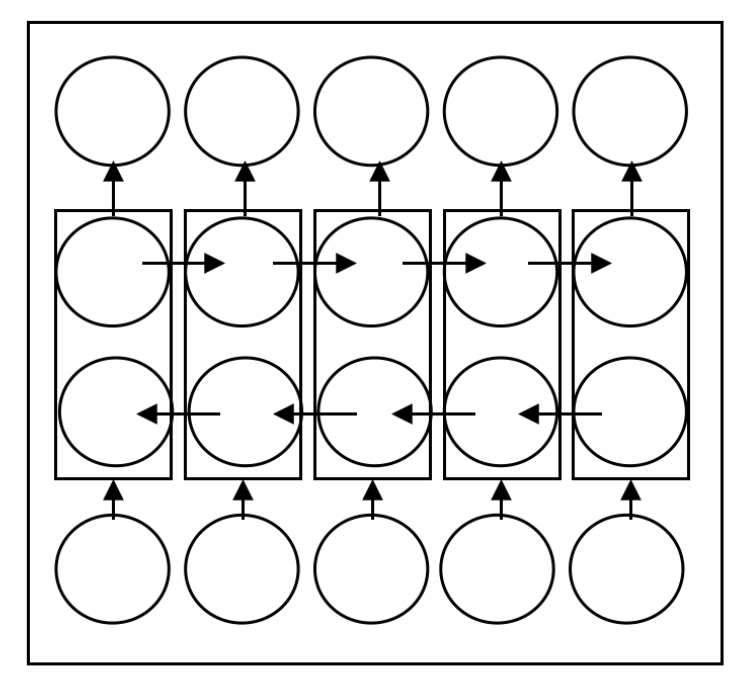}
\caption{Bidirectional RNNs}
\label{BRNN}
\end{figure}

Figure \ref{BRNN} illustrates the typical architecture of bidirectional RNNs. Notably, the hidden states are passed both from the initial to the final time steps and vice versa. This dual passage facilitates the formation of forward and backward connections, enhancing the network's understanding of the data sequence.

\textbf{Model Architecture}: BRNNs share the same architecture as RNNs except it is bidirectional rather than mono-directional.

\begin{figure}[h]
\centering
\includegraphics[width=15cm]{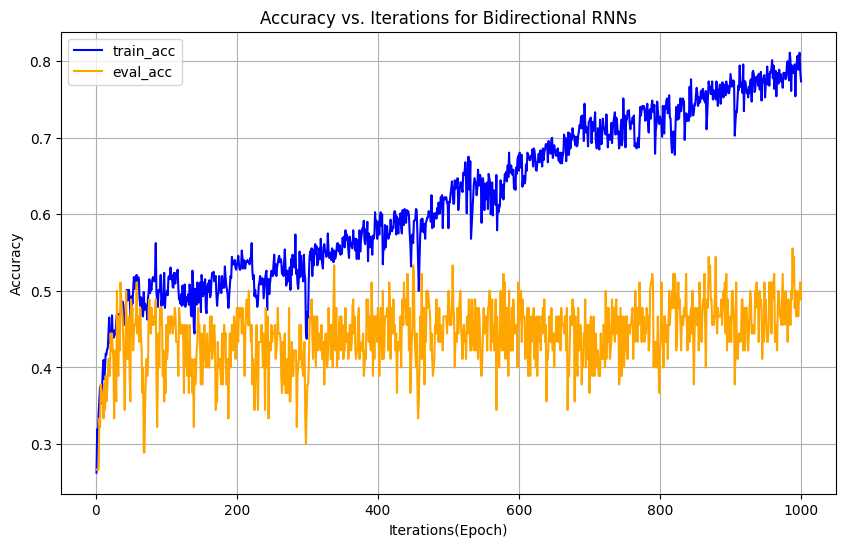}
\caption{Train and Evaluation Accuracy vs. Iterations for BRNNs}
\label{BRNNTestAcc}
\end{figure}

\textbf{Train}: We run 1000 epochs and plot the train accuracy and evaluation accuracy as shown in Figure \ref{BRNNTestAcc}.

After monitoring the training and evaluation accuracy across 10,000 epochs, we observed that the evaluation accuracy plateaued after 1,000 epochs. Consequently, we decided to halt the training at this point.

\subsubsection{LSTM}

Long Short-Term Memory (LSTM) networks are designed to address issues of gradient explosion and vanishing. They have the capability to forget non-essential information and retain important content. \\

\begin{figure}[h]
\centering
\begin{subfigure}{.5\textwidth}
  \centering
  \includegraphics[width=.8\linewidth]{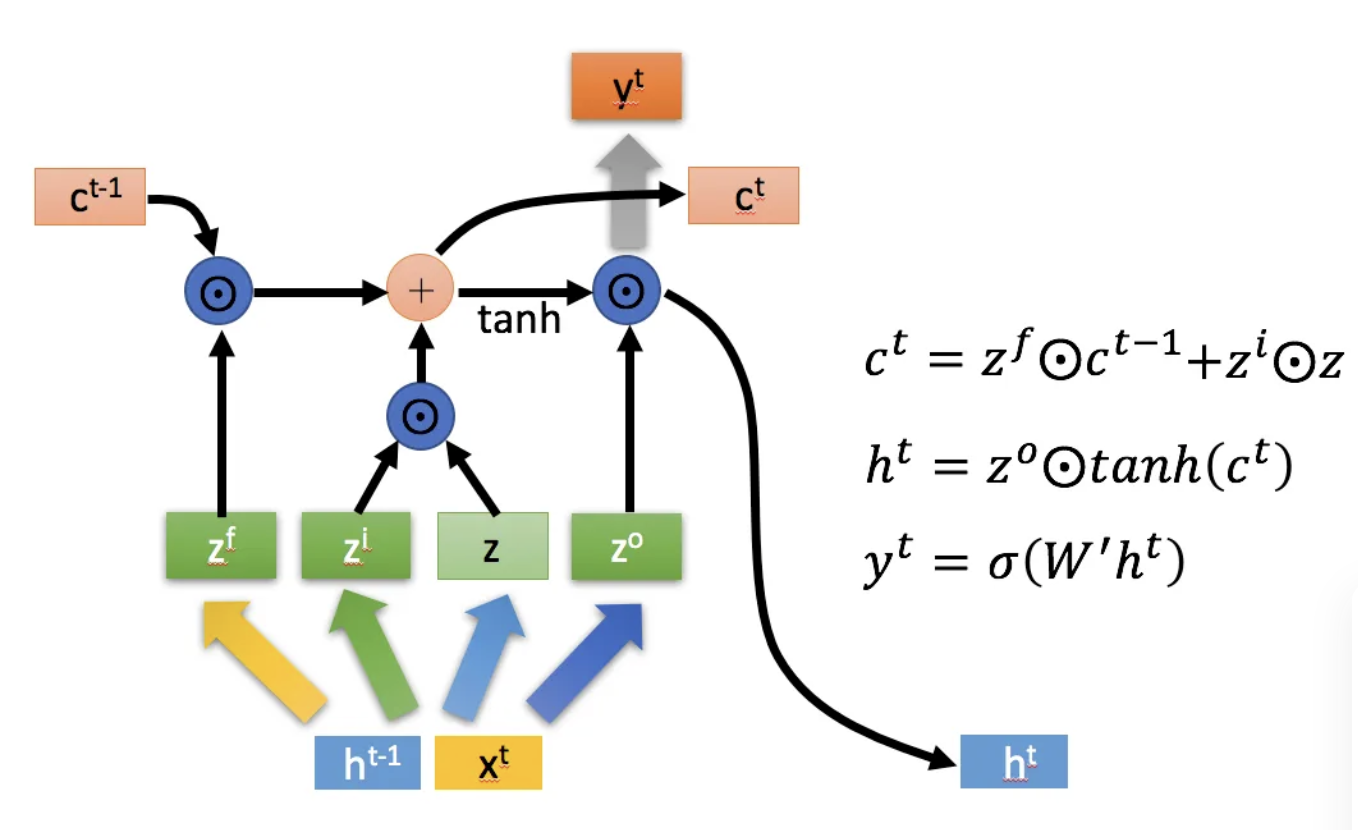}
  \caption{LSTM Model}
  \label{fig:sub1}
\end{subfigure}%
\begin{subfigure}{.5\textwidth}
  \centering
  \includegraphics[width=.5\linewidth]{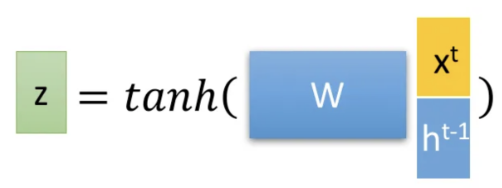}
  \caption{Formula of z in LSTM model}
  \label{fig:sub2}
\end{subfigure}
\caption{LSTM}
\label{fig:LSTM}
\end{figure}

Figure \ref{fig:sub1} depicts the structure of an LSTM model. Initially, the $z^f$ gate controls the aspects of the previous state,  $c^{t-1}$, that should be forgotten, selectively erasing unimportant elements. Next, the $z^i$ gate determines which new information to retain. This selected data, combined with the input from the previous state, $z$, and the result of the “forgetting” process, is passed to the next stage, $c^t$. Lastly, the zo gate manages the output $y^t$, controlling what is ultimately transmitted to the next layer or used for predictions.

\textbf{Model Architecture}: For LSTM, we construct 4 fully connected layers to make the model more complex so that it can help learn more details about our data. The first three layers are followed by batch normalization, RELU activation, and dropout. The final linear layer shapes the final output for the classification.

\begin{figure}[h]
\centering
\includegraphics[width=15cm]{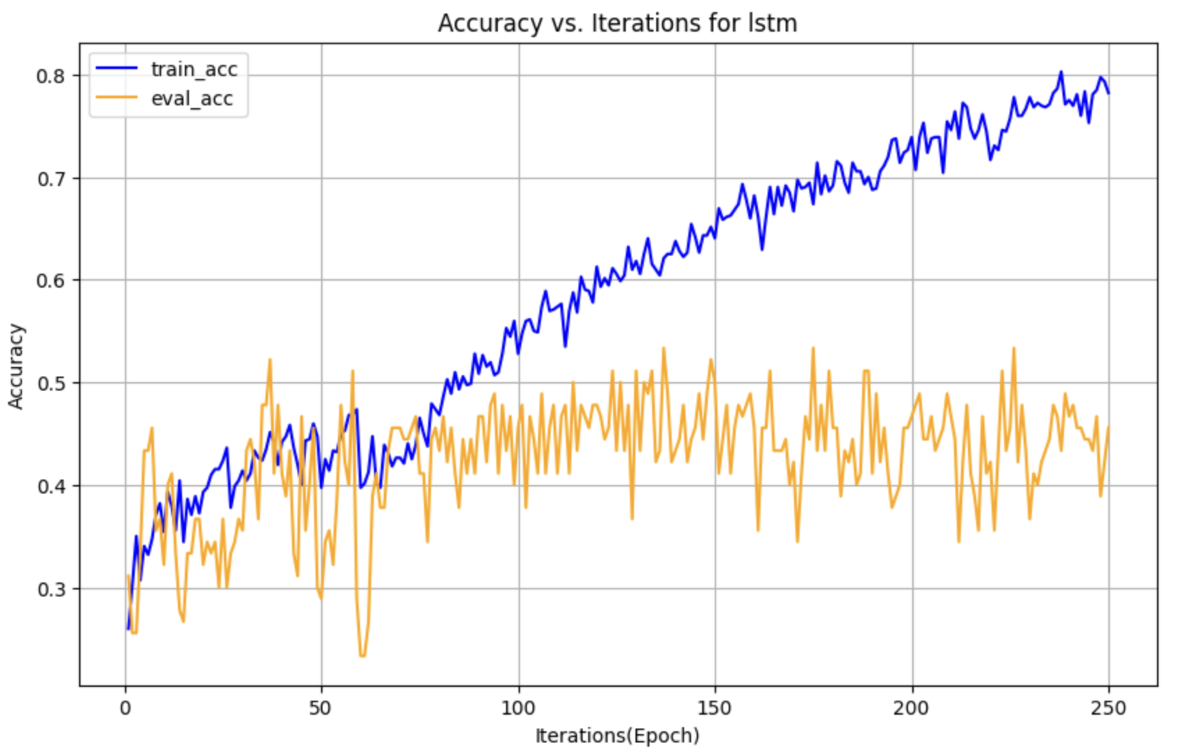}
\caption{Train and Evaluation Accuracy vs. Iterations for LSTM}
\label{LSTMTestAcc}
\end{figure}

\textbf{Train}: We run 250 epochs and plot the train accuracy and evaluation accuracy as shown in Figure \ref{LSTMTestAcc}.

We limited training to 250 epochs because we observed that the evaluation accuracy showed no significant changes beyond this point. However, it is worth noting that the training accuracy continued to exhibit a positive upward trend.

\subsection{Conclusions}

Table \ref{tab:table1} compares the test accuracy of three RNNs-related models: RNNs, BRNNs, and LSTM. Surprisingly, RNNs achieved the highest test accuracy at 53.33\%. Initially, we anticipated that LSTM would perform the best. However, the results shown in Table 1 may be influenced by our small datasets, which contain only 900 samples in total—720 for training and 90 each for evaluation and testing.

\begin{table}[h]
	\caption{Test Accuracy of RNNs, BRNNs, LSTM}
	\centering
	\begin{tabular}{lll}
		\toprule
		RNNs-Related Models     & Test Accuracy        \\
		\midrule
		RNNs      & 53.33\%      \\
		BRNNs     & 48.89\%   \\
		LSTM      & 51.11\%        \\
		\bottomrule
	\end{tabular}
	\label{tab:table1}
\end{table}

Given that RNNs are the simplest model among the three, this outcome suggests that simpler models often have better generalization capabilities, particularly with small datasets. Although the results were unexpected, the positive trend in training accuracy indicates that our models are performing well. We plan to test these models with a much larger dataset to see if they align more closely with our expectations.

\subsubsection{Baseline Models vs Neural Networks }

Comparing the neural networks and baseline models, the neural network models generally perform no better than the traditional machine learning models. This outcome may be attributed to the constrained size of the dataset. Simpler models are potentially more effective in extracting information from the dataset, whereas neural networks could lead to overfitting. Larger datasets will further be considered to test our model. 

\newpage
\section{Further Works 1 - Augment Original Data}

We suspect that the low model performance is due to the limited size of our dataset. To address this issue, we decide to augment the original dataset. There are many options to enlarge the dataset, we will use the techniques of adding random noise, time shifting, and pitch changing. These modifications will be applied to the original data to create a more robust dataset. Specifically, we expect to see positive outcomes with the expanded dataset, which will comprise 3600 data samples, each featuring 14 attributes and one label column. This augmented dataset should provide a more comprehensive basis for training our model, potentially leading to more accurate and reliable results.

\subsection{Dataset}

To enhance our dataset and improve the performance of our neural network, we have employed three different augmentation techniques on the original set of 900 samples. As a result, our dataset now includes a total of 3600 samples, with each of the four emotional quadrants—Q1, Q2, Q3, and Q4—equally represented by 900 samples.

To prepare this augmented data for neural network training, we have standardized the dataset format. Specifically, we padded and reshaped the data to ensure uniformity across samples. The resulting dataset is structured into an array of shapes (3600, 204, 1290), where 3600 represents the total number of samples, 204 is the number of features per sample, and 1290 denotes the number of time steps per sample. This uniform structuring is critical for training our neural network effectively, as it ensures that each input is consistently formatted and fully compatible with the network architecture.

\subsection{Models}

\subsubsection{Baseline}

\begin{figure}[h]
\centering
\includegraphics[width=15cm]{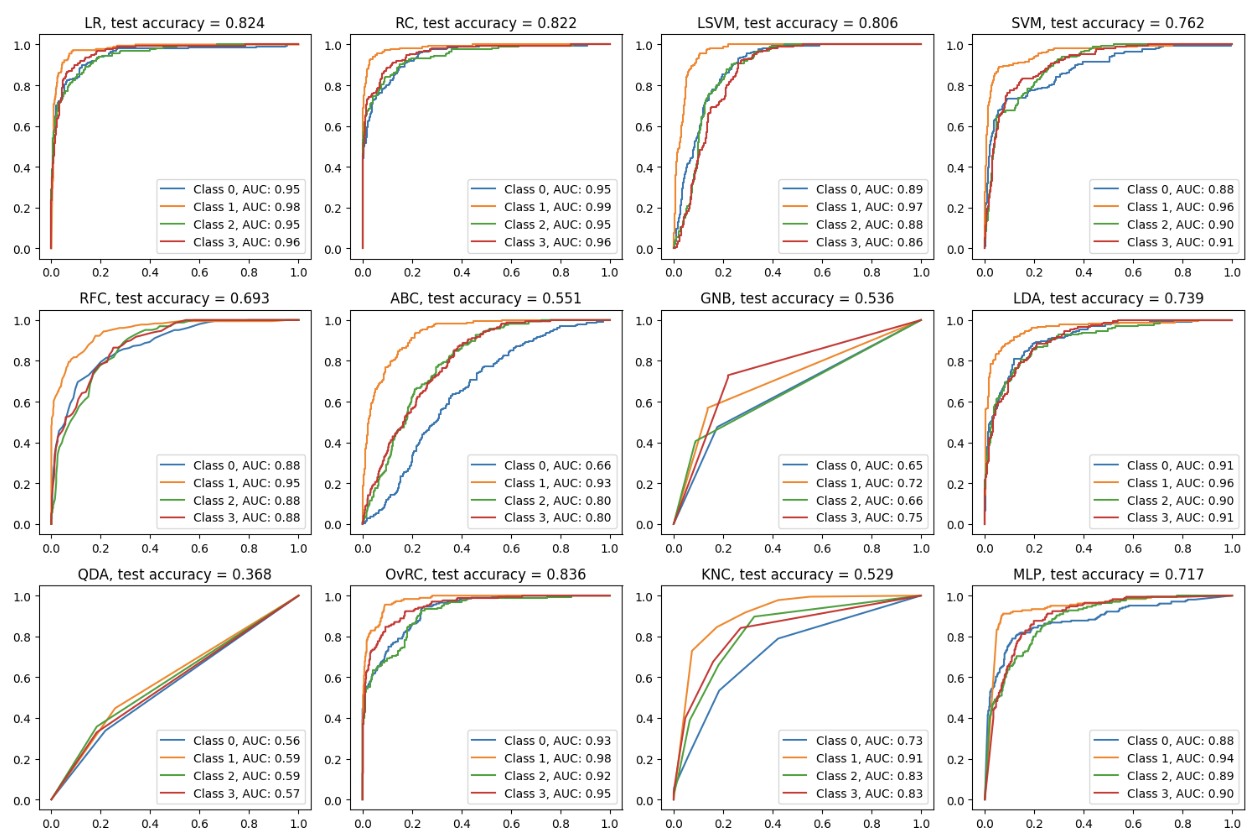}
\caption{Plots of Test Accuracy of Baseline Models on Augmented Dataset}
\label{BaselinePlotAug}
\end{figure}

We tested our baseline models on the augmented dataset. From the result of 12 scikit-learn classifiers shown in Figure \ref{BaselinePlotAug}, the top three best models are One-vs-Rest Classifier, Logistic Regression, and RidgeClassifier with accuracy 0.836, 0.824, 0.822 respectively. Compared with the original 4Q audio emotion dataset, test accuracy has generally improved, and it has increased by 20\% on models with good performance.

\subsubsection{RNN-Related Models}

To ensure the robustness and generalizability of our models, we implemented a 5-fold cross-validation approach during training for all three models: RNNs, BRNNs, and LSTM.

In this process, the dataset was divided into five distinct subsets. For each fold, one subset was used as the validation set while the remaining four were used for training. To prevent data leakage and ensure that the models generalize well to new data, we normalized the data in each fold using the statistical parameters (mean and standard deviation) calculated from the training subsets only.

Throughout the training process, we tracked and recorded the average validation loss for each fold. This measure helps us assess the performance and stability of the models across different subsets of the data, providing insights into their effectiveness and potential areas for improvement. 

Different from our original experiment, we used early-stopping techniques at this time. This is aimed to prevent over-training, which might negatively affect the results.

\begin{figure}[h]
\centering
\begin{subfigure}{.5\textwidth}
  \centering
  \includegraphics[width=1.1\linewidth]{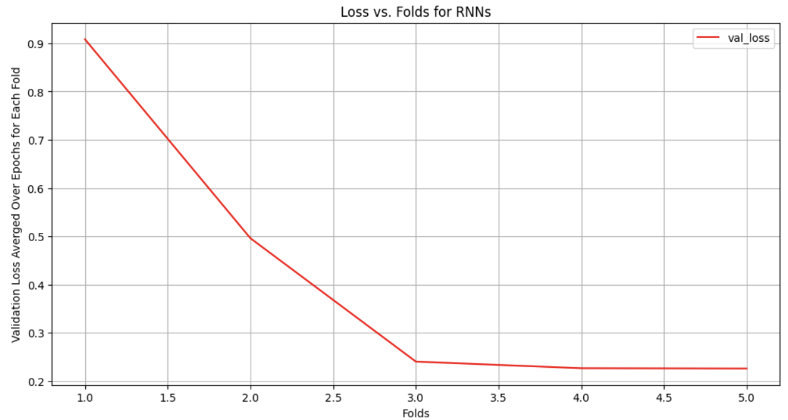}
  \caption{RNN}
  \label{fig:subRNN}
\end{subfigure}\\ 
\begin{subfigure}{.5\textwidth}
  \centering
  \includegraphics[width=1.1\linewidth]{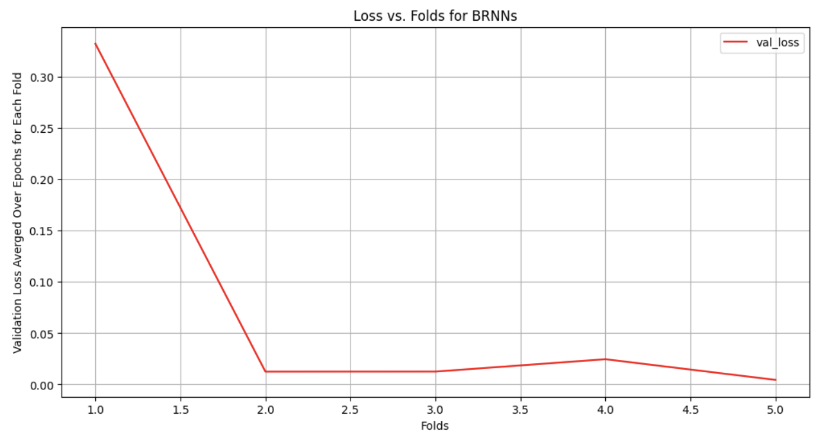}
  \caption{BRNN}
  \label{fig:subBRNN}
\end{subfigure}\\ 
\begin{subfigure}{.5\textwidth}
  \centering
  \includegraphics[width=1.1\linewidth]{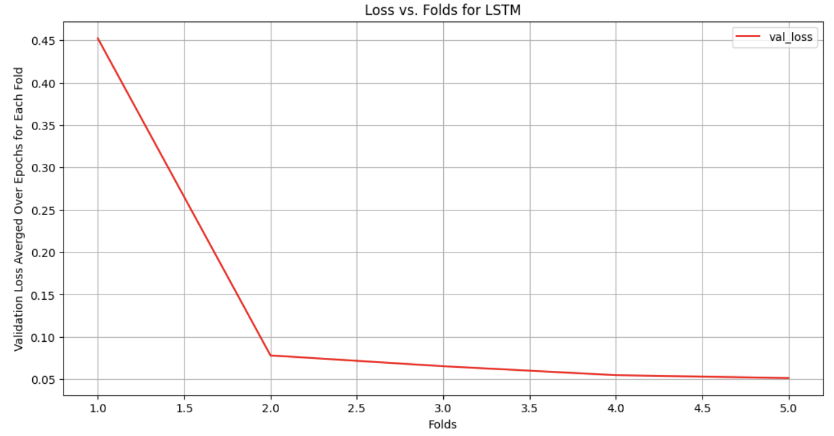}
  \caption{LSTM}
  \label{fig:subLSTM}
\end{subfigure}
\caption{Evaluation Loss vs. Folds}
\label{fig:EvalFold}
\end{figure}

Through the three plots about evaluation loss versus folds numbers shown in Figure \ref{fig:EvalFold}, we can notice that reaching the 5th fold, BRNNs show the lowest evaluation loss, then LSTM, then RNNs. From Table \ref{tab:table2} below, we can notice that the test performance for all three models improved. From the original data to augmented data, the test accuracy improved by 20\%, 19\%, and 30\% for RNNs, BRNNs, and LSTM, respectively. We fixed our model while applying the modifications to our dataset.

In our analysis, represented through three plots of evaluation loss versus fold numbers, the BRNNs demonstrated the lowest evaluation loss by the 5th fold, followed by LSTM, and RNNs. As shown in Table \ref{tab:table2}, transitioning from the original to the augmented dataset significantly improved test accuracy: RNNs by 20\%, BRNNs by 19\%, and LSTMs by 30\%. Throughout this evaluation, our model architecture remained constant, highlighting that improvements were solely due to the augmented dataset's enhanced quality and diversity. This confirms that dataset augmentation effectively boosts model performance.

\begin{table}[h]
	\caption{Test Accuracy of RNNs, BRNNs, LSTM between 900 and Augmented Data}
	\centering
	\begin{tabular}{lll}
		\toprule
		RNNs-Related Models & Test Accuracy(Original) & Test Accuracy(Augmented)       \\
		\midrule
		RNNs  & 53.33\%  &  64.03\% ↑\\
		BRNNs & 48.89\%  &  58.33\% ↑\\
		LSTM  & 51.11\%  &  65.97\% ↑\\
		\bottomrule
	\end{tabular}
	\label{tab:table2}
\end{table}

\subsection{Conclusion}

Comparing the neural networks and baseline models, the neural network models generally perform no better than the traditional machine learning models. However, the effect of data augmentation is very significant. The test accuracy increased by more than 10\% on all three neural networks. 

\newpage
\section{Further Works 2 - Larger Dataset}

We intend to implement the same baseline models alongside Recurrent Neural Networks (RNNs), Bidirectional Recurrent Neural Networks (BRNNs), and Long Short-Term Memory networks (LSTM) on larger datasets. Additionally, we aim to classify the dataset into more granular categories.  

\subsection{Datasets}
\begin{figure}[h]
\centering
\includegraphics[width=15cm]{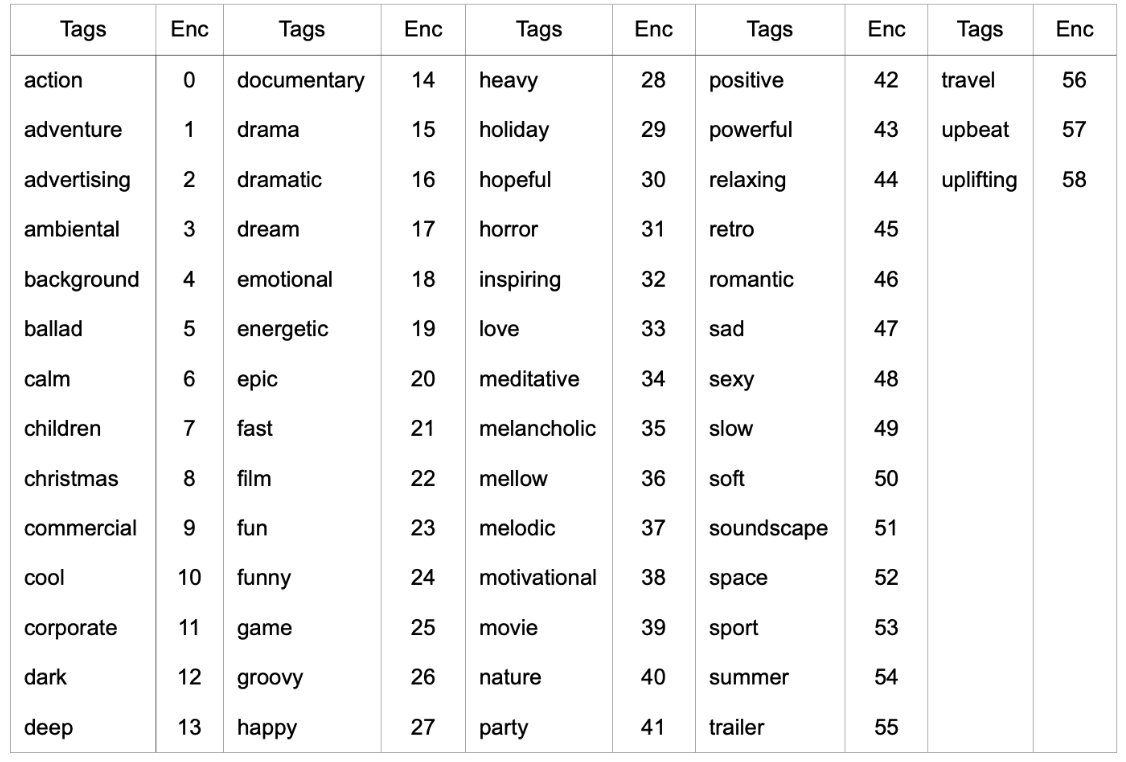}
\caption{59 Mood Genres and Their Encodings}
\label{59labels}
\end{figure}

We utilized the MTG-Jamendo Dataset\cite{bogdanov2019mtg}, a recently released open dataset for music auto-tagging which is accessible on GitHub. Given our limited computational resources, we opted for a subset of the dataset named “autotagging-moodtheme/audio-low,” comprising low-quality audio files specific to mood and theme. This subset includes approximately 14,000 audio tracks, each varying in length. The dataset is categorized into 59s distinct sub-genres, as outlined in Figure \ref{59labels}. These sub-genres serve as labels for training our model. Upon completion of the training process, we will manually map these sub-genre labels onto the four quadrants of Russell's Emotion Quadrant to interpret our model's output in terms of emotional content.

\subsection{Models}

\subsubsection{Baselines}

\begin{figure}[h]
\centering
\includegraphics[width=12cm]{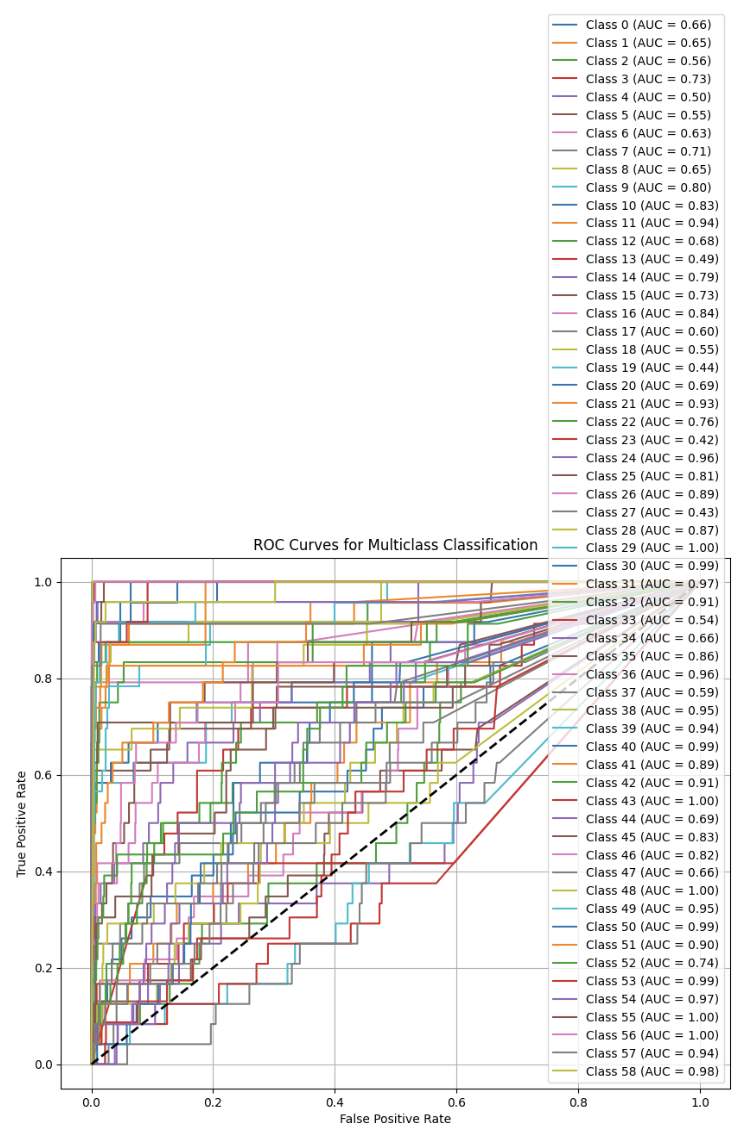}
\caption{ROC Curves For Multiclass Classification Through LR}
\label{59labelsROC}
\end{figure}

We applied a Logistic Regression model as the baseline model for the 59 classes classification and used AOC-ROC plot as Figure \ref{59labelsROC} to show the result. The model achieves 43.75\% accuracy on the test set.

\newpage

\newpage

\subsubsection{RNNs-Related Models}

\begin{figure}[h]
\centering
\begin{subfigure}{.5\textwidth}
  \centering
  \includegraphics[width=9cm]{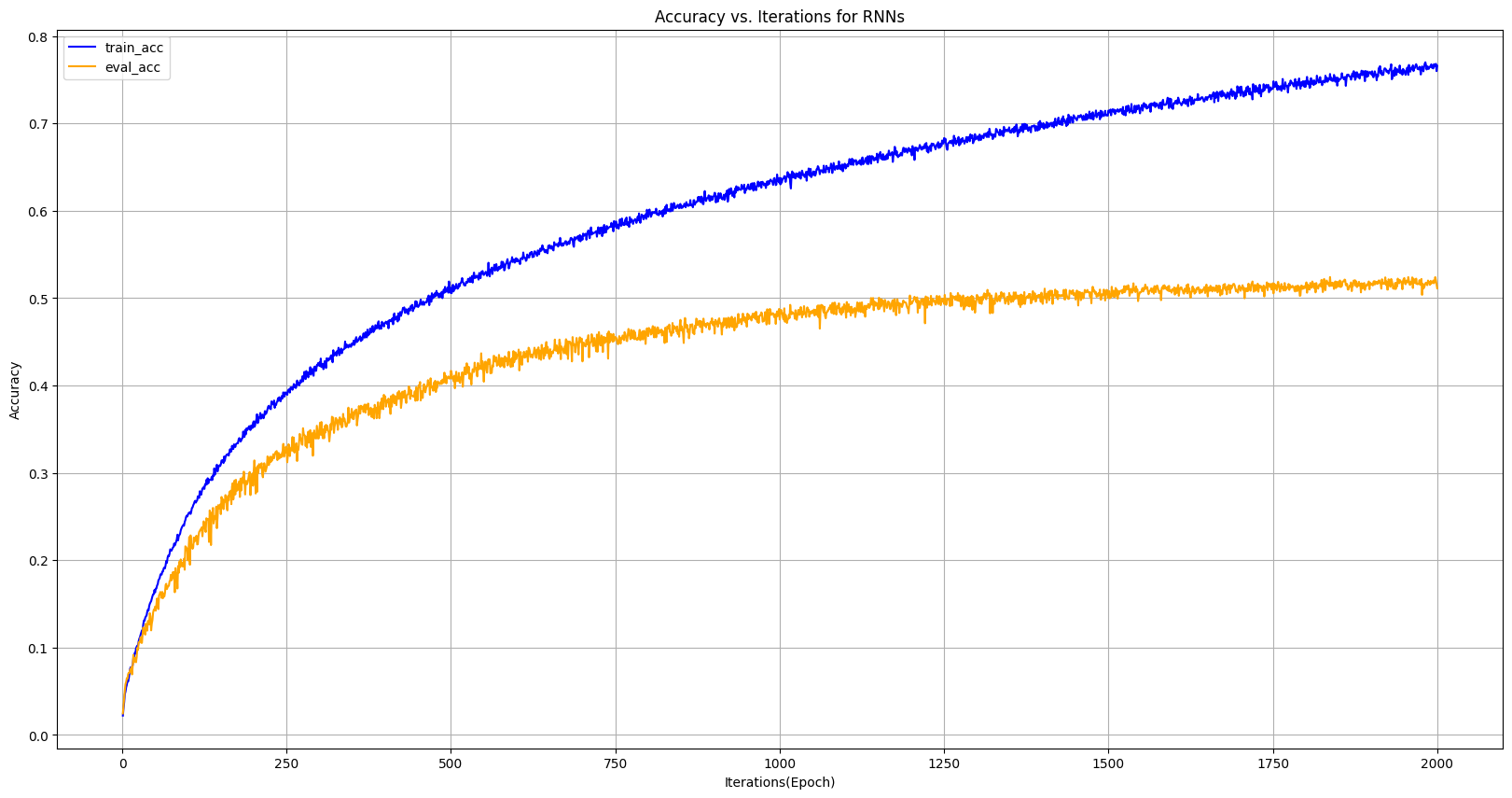}
  \caption{RNN}
  \label{fig:subRNNlarge}
\end{subfigure}\\ 
\begin{subfigure}{.5\textwidth}
  \centering
  \includegraphics[width=9cm]{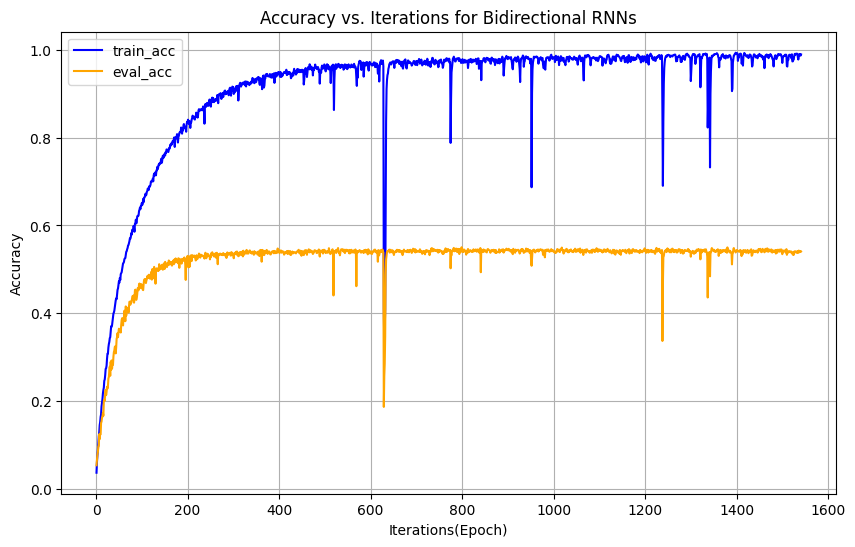}
  \caption{BRNN}
  \label{fig:subBRNNlarge}
\end{subfigure}\\ 
\begin{subfigure}{.5\textwidth}
  \centering
  \includegraphics[width=9cm]{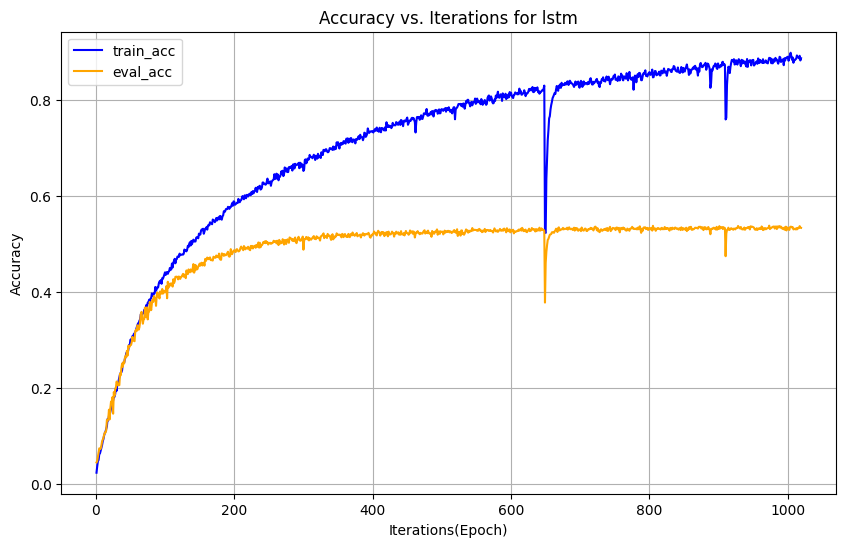}
  \caption{LSTM}
  \label{fig:subLSTMlarge}
\end{subfigure}
\caption{Train and Evaluation Accuracy vs. Iterations}
\label{fig:EvalLarge}
\end{figure}

Figure \ref{fig:subRNNlarge} displays a plot of training and evaluation accuracy versus 2,000 epochs trained for RNNs model. After the evaluation accuracy passed 0.5, the growth shown on the curve starts to stop.

From Figure \ref{fig:subBRNNlarge} we can notice that for BRNNs, 400 epochs is a fairly reasonable number to stop training. Though after 400 iterations, the training accuracy still increases, the evaluation accuracy is a line after that showing no tendency to increase anymore.

LSTM shows a similar situation as BRNNs. We can notice that the evaluation accuracy stops climbing after 400 iterations from Figure \ref{fig:subLSTMlarge}. 

\subsection{Results}
Table 2 shows a comparison of the test accuracy on three RNN-related models between our original 900 samples dataset and later 14,000 samples larger dataset.

\begin{table}[h]
	\caption{Test Accuracy of RNNs, BRNNs, LSTM between 900 and 14,000 Data}
	\centering
	\begin{tabular}{lll}
		\toprule
		RNNs-Related Models & Test Accuracy (900) & Test Accuracy (14,000)        \\
		\midrule
		RNNs      & 53.33\%  & 50.18\% $\downarrow$   \\
		BRNNs     & 48.89\%  & 53.97\% $\uparrow$\\
		LSTM      & 51.11\%  & 52.61\% $\uparrow$  \\
		\bottomrule
	\end{tabular}
	\label{tab:table}
\end{table}

From Table \ref{tab:table} we can clearly see the test accuracy of RNNs decreased as the dataset becomes larger, whereas both Bidirectional RNNs and LSTM have an increased test accuracy as we have more data samples for training the models. For a larger dataset, we have BRNNs as the model with the best performance among the three, where LSTM leads the second place, and lastly the RNNs model. This proved our hypothesis that larger dataset performs better on more complex models, whereas RNNs cannot capture many complex details within larger datasets.

\subsection{Conclusions}

Comparing the result of Logistic Regression and neural networks, our model performs significantly better than logistic regression, which indicates that traditional models may be more suitable for small and medium-size datasets. Neural Network models are better suited to capture the complex information of large data sets. 

When comparing the results across datasets of varying sizes, we observed that RNNs, BRNNs, and LSTMs performed best on the medium-sized dataset containing 3,600 audio clips. This outcome may be attributed to the fact that larger datasets necessitate more complex models (either deeper or with more intricate structures) to effectively capture information. Our current model is best suited for medium-sized datasets.

\subsection{Other Notes}

\textit{Whether for the larger dataset, a more complex model shows a better performance?}

Within the three RNN models, in the result section of further works 1, LSTM do have the best test accuracy among the three with the most complex model of the three. However, some traditional machine learning models are still outperforming the three RNN models. It is worth testing whether a deep neural network will perform better in the future.

\textit{Whether RNN-related models are good for music emotion classification?}

From our comparison, RNN-related models perform no better than the traditional machine learning models.

\textit{What are the possible biases in this project?}

\textbf{Data bias}: only takes about the first 30 seconds’ features into training due to limited resources

\textbf{Labeling bias}: cannot manually change 59 mood genres into 4 quadrants, therefore using 59 labels to train the large dataset and record the results in Further Works 2.

There are still many aspects worth improving in our project for future work.


\bibliographystyle{plainnat} 






\end{document}